\documentstyle[amssymb,12pt]{article}
\begin{document}

\title{Asymptotic Estimation of Shift Parameter of a Quantum State\thanks{
Work partially supported by INTAS grant 00-738.}}
\author{A. S. Holevo}
\date{}
\maketitle

\section{\protect\bigskip Introduction}

The rise of quantum estimation theory dates back to the late 1960s -- 1970s
(see \cite{hel76}, \cite{hol80}). It was initially developed as an adequate
mathematical framework for design of optimal receiver in quantum
communication channels, and later turned out to be relevant also for
clarification of some foundational issues of quantum measurement. New
interest to quantum estimation theory was brought by the development of high
precision experiments, in which researchers operate with elementary quantum
systems. In such experiments quite important is the issue of extracting the
maximum possible information from the state of a given quantum system. For
example, in currently discussed proposals for quantum computations, the
information is written into states of elementary quantum cells -- qubits,
and is read off via quantum measurements. From a statistical viewpoint,
measurement gives an estimate for the quantum state, either as a whole, or
for some of its components (parameters). The main concrete models of
interest in quantum estimation theory considered up to now fall within one
of the following classes:

1) Parametric models with a group of symmetries \cite{hol80}, \cite{2*}. In
particular, the models with the shift or rotation parameter are strictly
relevant to the issue of canonical conjugacy and nonstandard uncertainty
relations, such as time-energy, phase-number of quanta, etc. It is these
models which will be our main concern in this paper.

2) The full model, in which the multidimensional parameter is the quantum
state itself, i. e. we are interested in estimation of completely unknown
quantum state. Although in finite dimensions it is a parametric model with a
specific group of symmetries, it deserves to be singled out both because of
its importance for physics and of its mathematical features. Especially
interesting and mostly studied is the case of the qubit state, with the
3-dimensional parameter varying inside the Bloch sphere. Asymptotic
estimation theory for the full model in the pure state case was developed in
\cite{ma}, \cite{hay}, and for mixed states in \cite{key}.

On the other hand, the full model, especially in infinite dimensions,
belongs rather to nonparametric quantum mathematical statistics, which is at
present also in a stage of development. In this connection we would like to
mention the method of homodyne tomography of a density operator in quantum
optics \cite{7*}, \cite{dar}.

3) Estimation of the mean value of quantum Gaussian states. This is a
quantum analog of the classical linear ``signal+noise'' problems, however
with the noise having quantum-mechanical origin. This model was treated in
detail e.g. in \cite{hol80}.

An important distinctive feature of quantum estimation appears in
consideration of series of independent identical trials of a quantum system.
In the paper \cite{hol83}, devoted to asymptotics of estimation of a shift
parameter in a quantum state, it was observed that {\it entangled }
covariant estimates in models with independent multiple observations can be
more efficient than unentangled ones. For arbitrary locally unbiased
observables this was demonstrated on the full model in \cite{ma1}, \cite{ma},
\cite{hay}, \cite{der}. This property is a statistical counterpart of the
strict superadditivity of Shannon information for quantum memoryless channels
(see.  \S 5.1 in \cite{2*}).

In the present paper we develop further the asymptotic theory of estimation
of a shift parameter in a quantum state to demonstrate the relation between
entangled and unentangled covariant estimates in the analytically most
transparent way. After recollecting basics of estimation of shift parameter
in Sec. 2, we study the structure of the optimal covariant estimate in Sec.
3, showing how entanglement comes into play for several independent trials.
In Secs. 4,5 we give the asymptotics of the performance of the optimal
covariant estimate comparing it with the ``semiclassical''\ unentangled
covariant estimation in the regular case of finite variance of the generator
of the shift group. Sec. 6 is devoted to estimation in the case where the
regularity assumption is violated.

In this paper we deal with the case of pure states. It is in fact yet
another distinctive feature of quantum estimation that the complexity of the
problem increases sharply with transition from pure to mixed states
estimation. In fact, estimation theory for mixed states is an important
field to a great extent still open for investigation. Another simplifying
factor in our model is that the parameter is one-dimensional. It is well
known that in the quantum case estimation of multidimensional parameter
involves additional problems due to the non-commutativity of the algebra of
quantum observables, see e. g. \cite{hol80}.

\section{Covariant Estimates}

Let ${\cal H}$ be a Hilbert space of observed quantum system. The states of
the system are described by density operators $S$ in ${\cal H},$ and
observables with values in a measurable space ${\cal X}$ -- by resolutions
of the identity or probability operator valued measures $M$ on ${\cal X}$
(see, e. g. \cite{hol80}, \cite{2*}). Let $G$ be either the real line ${\Bbb
R}$ (the case of a displacement parameter) or the unit circle ${\Bbb T}$
(the case of a rotational parameter), and let $x\rightarrow V_{x}=e^{-ixA};$
$x\in G,$ be a unitary representation of the one dimensional Abelian group $
G $ in ${\cal H}$. The spectrum $\Lambda $ of the operator $A$ is contained
in the dual group $\widehat{G}$, which is identified with ${\Bbb R}$ in the
case $G={\Bbb R}$ and with the group of integers ${\Bbb Z}$ when $G={\Bbb T}$
. We consider the problem of estimation of the shift parameter $\theta \in G$
in the family of states
\[
S_{\theta }=e^{-i\theta A}Se^{i\theta A},\quad
\]
where $S$ is the inital state, assumed to be known. By {\it estimate} we
call arbitrary observable with values in $G.$

The estimate $M$ is {\it covariant} if

\begin{equation}
V_{x}^{\ast }M(B)V_{x}=M(B-x)\,\quad {\rm for\quad }\,B\subset G,x\in G,
\label{H2.3.9}
\end{equation}
where $B-x:=\{y;y+x\in B\}$, and in the case $G={\Bbb T}$ addition is modulo
$2\pi $. {\it A necessary and sufficient condition for existence of a
covariant observable is that the spectrum of }$A${\it \ is absolutely
continuous with respect to the Haar measure in the dual group} $\widehat{G}$
\cite{hol2}, which we assume from now on. In the case $G={\Bbb R}$ this
means that $A$ has Lebesgue spectrum, while in the case $G={\Bbb T}$ (with $
\widehat{G}={\Bbb Z}$) this poses no restrictions. We introduce the operators

\begin{equation}
U_{y}:=\int_{G}{\rm e}^{iyx}M(dx);\quad y\in \widehat{G.}  \label{H2.3.19a}
\end{equation}
Then (\ref{H2.3.9}) reduces to the Weyl relation

\begin{equation}
U_{y}V_{x}={\rm e}^{ixy}V_{x}U_{y};\quad \,\,x\in G,y\in \widehat{G},
\label{2.3.10}
\end{equation}
in which, however, the operator $U_{y}$, is in general non-unitary. In this
sense the observable $M$ is canonically conjugate to the observable $A$.

Introducing the characteristic function
\[
\varphi _{S}^{M}(\lambda )={\sf E}_{S}U_{\lambda }=\int_{G}{\rm e}^{i\lambda
x}\mu _{S}^{M}(dx),\quad \lambda \in \widehat{G},
\]
one has the following uncertainty relation for the generalized canonical
pair $(A,M)$ \cite{hol83}

\begin{equation}
\Delta _{S}^{M}(\lambda )\cdot {\sf D}_{S}({A)}\geq \frac{1}{4}
\label{H2.3.11}
\end{equation}
where $\Delta _{S}^{M}(\lambda ):=\lambda ^{-2}(|\varphi _{S}^{M}(\lambda
)|^{-2}-1)$, $\lambda \neq 0$, is a functional measure of uncertainty of the
covariant observable $M$ in the state $S$. If $G={\Bbb R}$ and $M$ has
finite variance ${\sf D}_{S}{(M)}$, then $\lim_{\lambda \rightarrow 0}\Delta
_{S}^{M}(\lambda )={\sf D}_{S}{(M)}$, so that from (\ref{H2.3.11}) follows
the generalization of the Heisenberg uncertainty relation

\begin{equation}
{\sf D}_{S}{(}M{)}\cdot {\sf D}_{S}(A{)}\geq \frac{1}{4}.  \label{unc}
\end{equation}

Of the main interest are covariant estimates having the minimal uncertainty.
The following result describes them in the case of a pure initial state $S=$
$|\psi {\rangle \langle }\psi {|,}$ where $\psi $ is a unit state vector in $
{\cal H}$. Then
\[
S_{\theta }={\rm e}^{-i\theta A}|\psi {\rangle \langle }\psi {|}{\rm e}
^{i\theta A}.
\]
Since the spectrum of $A$ is absolutely continuous, we have the direct
integral spectral decomposition
\begin{equation}
{\cal H}=\int_{\Lambda }\oplus {\cal H}(\lambda )d\lambda ,  \label{di}
\end{equation}
diagonalizing the unitary group$\left\{ {\rm e}^{i\theta A}\right\} $. This
means that for any vector
\[
|\varphi {\rangle }=\int_{\Lambda }\oplus |\varphi (\lambda ){\rangle }
d\lambda ,\quad |\varphi (\lambda ){\rangle \in }{\cal H}(\lambda ),
\]
one has
\[
e^{i\theta A}|\varphi {\rangle }=\int_{\Lambda }\oplus e^{i\theta \lambda
}|\varphi (\lambda ){\rangle }d\lambda ,\quad \theta \in G.
\]
We agree to extend $\varphi (\lambda )$ to the whole of $\hat{G}$ by letting
it zero outside of $\Lambda .$

{\bf Theorem} {\bf 1 }\cite{hol83}. {\it For arbitrary covariant estimate }$
M $
\begin{equation}
\left| \varphi _{S}^{M}(\lambda )\right| \leq \varphi _{S}^{\ast }(\lambda
):=\int_{\hat{G}}\left\| \psi (\lambda ^{\prime })\right\| \left\| \psi
(\lambda ^{\prime }+\lambda )\right\| d\lambda ^{\prime },  \label{chf}
\end{equation}
\begin{equation}
\Delta _{S}^{M}(\lambda )\geq \Delta _{S}^{\ast }(\lambda ):=\lambda
^{-2}\left( \varphi _{S}^{\ast }(\lambda )^{-2}-1\right) .  \label{H2.3.12}
\end{equation}
{\it In the case} $G={\Bbb R}$
\begin{equation}
{\sf D}_{S}{(M)\geq }{\sf D}_{S}^{\ast }{:}=\int_{{\bf R}}\left( \frac{d}{
d\lambda }\left\| \psi (\lambda )\right\| \right) ^{2}d\lambda ,
\label{H2.3.13}
\end{equation}
{\it provided the right hand side is defined and finite.}

{\it The equalities are attained on the optimal covariant observable }$
M_{\ast }${\it \ which is given by the following kernel in the direct
integral decomposition (\ref{di})}
\begin{equation}
M_{\ast }(dx)=\left[ {\rm e}^{ix(\lambda ^{\prime }-\lambda )}\frac{|\psi
(\lambda ){\rangle }\,\,{\langle }\psi (\lambda ^{\prime }){|}}{\Vert \psi
(\lambda )\Vert \Vert \psi (\lambda ^{\prime })\Vert }\right] \frac{dx}{2\pi
}.\quad \Box  \label{m*}
\end{equation}

Notice that the operator
\begin{equation}
P_{\ast }:=\int_{G}M_{\ast }(dx)=\left[ \delta (\lambda ^{\prime }-\lambda )
\frac{|\psi (\lambda ){\rangle \langle }\psi (\lambda ^{\prime }){|}}{\Vert
\psi (\lambda )\Vert \Vert \psi (\lambda ^{\prime })\Vert }\right]
\label{p*}
\end{equation}
is a projection onto the invariant subspace of the group $\left\{ {\rm e}
^{i\theta A}\right\} $ generated by the vectors ${\rm e}^{i\theta A}|\psi {\
\rangle ,}\theta \in G{.}$ Indeed, $P_{\ast }{\rm e}^{i\theta A}|\psi {\
\rangle =}{\rm e}^{i\theta A}|\psi {\rangle }$ and if the vector $|\varphi {
\ \ \rangle }$ is orthogonal to all of these vectors, then $\int_{\Lambda }
{\rm e }^{i\lambda \theta }{\langle }\varphi (\lambda ){|}\psi (\lambda ){\
\rangle } d\lambda =0,$ $\theta \in G,$ and ${\langle }\varphi (\lambda ){|}
\psi (\lambda ){\rangle }=0$ for $\lambda \in \Lambda ,$ hence $P_{\ast
}|\varphi {\rangle =0.}$ Thus $M_{\ast }$ is in general subnormalized, and
one has to extend it to the orthogonal complement of ${\cal H}_{\ast
}=P_{\ast }{\cal H} $ to obtain an observable. Independently of the
extension, it has the following probability density in the state $S_{\theta
},$ given by the Fourier transform of the characteristic function (\ref{chf}
),
\[
p_{\theta }^{\ast }(x)=\frac{1}{2\pi }\left| \int {\rm e}^{i\lambda
(x-\theta )}\Vert \psi (\lambda )\Vert d\lambda \right| ^{2}.
\]

In \cite{hol2} it is shown also that $M_{\ast }$ {\it minimizes the average
deviation } ${\cal R}(M)=\int W(x-\theta )p_{\theta }^{M}(x)dx$ {\it where }
$W$ {\it is an arbitrary continuous conditionally negative definite
function. }

Similarly to (\ref{p*}), we obtain for future use
\begin{equation}
U_{\ast y}:=\int_{G}{\rm e}^{iyx}M_{\ast }(dx)=\left[ \delta (y+\lambda
^{\prime }-\lambda )\frac{|\psi (\lambda ){\rangle }\,\,{\langle }\psi
(\lambda ^{\prime }){|}}{\Vert \psi (\lambda )\Vert \Vert \psi (\lambda
^{\prime })\Vert }\right] .  \label{u*}
\end{equation}

{\bf Example} \cite{hel76}, \cite{hol80}. Consider the case ${\cal H=}L^{2}(
{\Bbb R}),$ where $A$ acts as multiplication by the independent variable $
\lambda $ (momentum representation). Then $\theta $ is the position
displacement parameter. In this case ${\cal H}(\lambda )\simeq {\Bbb R},$ $
\psi (\lambda )/\left\| \psi (\lambda )\right\| ={\rm e}^{i\alpha (\lambda
)} $ and (\ref{m*}) is an orthogonal resolution of the identity in ${\cal H}$
. This is most easily seen by verifying that the operators (\ref{u*}) form a
unitary group. This resolution of the identity is the spectral measure of
the selfadjoint operator
\begin{equation}
Q_{\ast }=\int xM_{\ast }(dx)=\frac{1}{i}\frac{d}{dy}|_{y=0}U_{\ast y}.
\label{q*}
\end{equation}
According to (\ref{u*}) the action of this operator on the vector $\varphi $
is given by
\[
Q_{\ast }\varphi (\lambda )=i{\rm e}^{i\alpha (\lambda )}\frac{d}{d\lambda }
{\rm e}^{-i\alpha (\lambda )}\varphi (\lambda ),
\]
where $Q=i\frac{d}{d\lambda }$ is the position operator in the momentum
representation. Thus the optimal covariant observable is position observable
up to the gauge transformation compensating the phase of the state vector $
|\psi {\rangle }$ in the momentum representation. In case the argument $
\alpha (\lambda )$ is absolutely continuous one has further
\[
Q_{\ast }\varphi (\lambda )=\left[ Q+\alpha ^{\prime }(\lambda )\right]
\varphi (\lambda )
\]
on an appropriate domain of functions $\varphi (\lambda ).$

\section{Multiple Observations}

Now we consider the problem of estimation of the shift parameter $\theta \in
{\Bbb R}$ in the family of states
\begin{equation}
S_{\theta }^{\otimes n}=S_{\theta }\otimes \dots \otimes S_{\theta }
\label{sn}
\end{equation}
in the Hilbert space ${\cal H}^{\otimes n}={\cal H}\otimes \dots \otimes
{\cal H}$ ($n$-fold tensor product which corresponds to $n$ independent
observations). Here $S_{\theta }={\rm e}^{-i\theta A}S{\rm e}^{i\theta A}$,
where $S=|\psi \rangle \langle \psi |$ is a pure state.

The family (\ref{sn}) is covariant with respect to the unitary
representation of the group of shifts of ${\Bbb R}$
\[
\theta \rightarrow \exp (-i\theta A^{(n)}),\quad A^{(n)}=A\otimes \dots
\otimes I+\dots +I\otimes \dots \otimes A
\]
in ${\cal H}^{\otimes n}$. The corresponding direct integral decomposition
reads
\[
{\cal H}^{\otimes n}=\int \oplus {\cal H}^{(n)}(\lambda )d\lambda ,
\]
where
\[
{\cal H}^{(n)}(\lambda )=\int \oplus \left[ {\cal H}(\lambda _{1})\otimes
\dots \otimes {\cal H}(\lambda _{n-1})\otimes {\cal H}(\lambda -\lambda
_{1}-\dots -\lambda _{n-1})\right] d\lambda _{1}\dots d\lambda _{n-1}.
\]
For symmetry of notations we shall also denote this as
\[
\int_{\Gamma (\lambda )}\oplus \left[ {\cal H}(\lambda _{1})\otimes \dots
\otimes {\cal H}(\lambda _{n})\right] d^{n-1}\sigma ,
\]
where
\[
\Gamma (\lambda )=\left\{ (\lambda _{1},\dots ,\lambda _{n}):\lambda
_{1}+\dots +\lambda _{n}=\lambda \right\} .
\]
Then $|\psi ^{\otimes n}\rangle =\int \oplus |\psi ^{(n)}(\lambda )\rangle
d\lambda ,$ where
\[
|\psi ^{(n)}(\lambda )\rangle =\int_{\Gamma (\lambda )}\oplus \left[ |\psi
(\lambda _{1})\rangle \otimes \dots \otimes |\psi (\lambda _{n})\rangle
\right] d^{n-1}\sigma ,
\]
so that
\begin{equation}
\left\| \psi ^{(n)}(\lambda )\right\| ^{2}=\int_{\Gamma (\lambda )}\oplus
\left[ \left\| \psi (\lambda _{1})\right\| ^{2}\otimes \dots \otimes \left\|
\psi (\lambda _{n})\right\| ^{2}\right] d^{n-1}\sigma .  \label{conv}
\end{equation}
With these relations in mind, the optimal covariant observable is given by
the formula (\ref{m*})
\begin{equation}
M_{\ast }^{(n)}(dx)=\left[ {\rm e}^{ix(\lambda ^{\prime }-\lambda )}\frac{
|\psi ^{(n)}(\lambda ){\rangle }\,{\langle }\psi ^{(n)}(\lambda ^{\prime }){
\ | }}{\Vert \psi ^{(n)}(\lambda )\Vert \Vert \psi ^{(n)}(\lambda ^{\prime
})\Vert }\right] \frac{dx}{2\pi }.  \label{m*n}
\end{equation}
However in this case the projection
\begin{equation}
P_{\ast }^{(n)}:=\int M_{\ast }^{(n)}(dx)=\left[ \delta (\lambda ^{\prime
}-\lambda )\frac{|\psi ^{(n)}(\lambda ){\rangle }\,{\langle }\psi
^{(n)}(\lambda ^{\prime }){|}}{\Vert \psi ^{(n)}(\lambda )\Vert \Vert \psi
^{(n)}(\lambda ^{\prime })\Vert }\right] ,\quad n>1,  \label{p*n}
\end{equation}
cannot be equal to the identity operator: in any case it projects onto a
subspace ${\cal H}_{\ast }^{(n)}$ lying in the subspace of vector functions
symmetrically depending on $\lambda _{1},\dots ,\lambda _{n}.$

Let us denote $p(\lambda )=\left\| \psi (\lambda )\right\| ^{2}$ and $
p^{(n)}(\lambda )=\left\| \psi ^{(n)}(\lambda )\right\| ^{2}$ the
probability densities of the observables $A$ and $A^{(n)}$ in the states $S$
and $S^{\otimes n}$ respectively. Then (\ref{conv}) implies that $
p^{(n)}(\lambda )$ is the $n-$th convolution of $p(\lambda ),$ which we
denote as $p^{(n)}(\lambda )=p(\lambda )^{\ast n}.$ We assume that $
p(\lambda )$ and hence $p^{(n)}(\lambda )$ are differentiable functions.
Especially useful in this context is the concept of weak differentiability
\cite{ba}: the probability density $p(\lambda )$ is {\it weakly
differentiable} if there exists a function $s(\cdot )\in L^{2}(p),$ such
that for all $f$ with $\int \left| f(\lambda )\right| ^{2}p(\lambda +\theta
)d\lambda <\infty $ the function $g(\theta )=\int f(\lambda )p(\lambda
+\theta )d\lambda $ has a derivative $g^{\prime }(\theta )=\int f(\lambda
)s(\lambda +\theta )p(\lambda +\theta )d\lambda .$

To get more insight into the structure of the optimal covariant observable
for $n>1,$ let us consider it in the situation of the Example. Then ${\cal H}
^{\otimes n}$ is isomorphic to $L^{2}({\Bbb R}^{n}),$ and $A^{(n)}$ is just
the operator of multiplication by $\lambda =\lambda _{1}+\dots +\lambda _{n}.
$

{\bf Theorem 2}. {\it The optimal covariant observable (\ref{m*n}) is the
spectral measure of the selfadjoint operator}
\begin{equation}
Q_{\ast }^{(n)}=P_{\ast }^{(n)}\frac{1}{n}\left( Q_{\ast }\otimes \dots
\otimes I+\dots +I\otimes \dots \otimes Q_{\ast }\right) P_{\ast }^{(n)}.
\label{sum}
\end{equation}

Observable $\frac{1}{n}\left( Q_{\ast }\otimes \dots \otimes I+\dots
+I\otimes \dots \otimes Q_{\ast }\right) $ corresponds to a
``semi-classical'' method of estimation, when the optimal quantum estimates
for each of $n$ components in ${\cal H}^{\otimes n}$ are found and then used
in a classical way to obtain the average over $n$ observations. The theorem
shows that projecting this average onto ${\cal H}_{\ast }^{(n)}$ (and thus
introducing entanglement) gives the optimal quantum estimate for $n${\it \ }
observations.

{\it Proof. }As in the Example, one shows that (\ref{m*n}) is orthogonal
resolution of the identity. As follows from (\ref{p*n}) the subspace ${\cal H
}_{\ast }^{(n)}=P_{\ast }^{(n)}{\cal H}$ consists of the functions of the
form
\begin{equation}
\varphi (\lambda _{1},\dots ,\lambda _{n})=\frac{c(\lambda )}{\sqrt{
p^{(n)}(\lambda )}}\psi (\lambda _{1})\dots \psi (\lambda _{n}),  \label{phi}
\end{equation}
where $\int \left| c(\lambda )\right| ^{2}d\lambda =\left\| \varphi \right\|
^{2}.$ Consider the action on these functions of the selfadjoint operator
\[
Q_{\ast }^{(n)}=\int xM_{\ast }^{(n)}(dx)=\left[ \int x\frac{{\rm e}
^{-ix\lambda }|\psi ^{(n)}(\lambda ){\rangle \langle }\psi ^{(n)}(\lambda
^{\prime }){|}{\rm e}^{ix\lambda ^{\prime }}}{\Vert \psi ^{(n)}(\lambda
)\Vert \Vert \psi ^{(n)}(\lambda ^{\prime })\Vert }\frac{dx}{2\pi }\right] .
\]
(Notice a parallel between this expression and the Pitman formula
\begin{equation}
\theta _{\ast }=\frac{\int \theta \tilde{p}(x_{1}-\theta )\dots \tilde{p}
(x_{n}-\theta )d\theta }{\int \tilde{p}(x_{1}-\theta )\dots \tilde{p}
(x_{n}-\theta )d\theta }  \label{pit}
\end{equation}
for the classical optimal covariant estimate of the shift parameter $\theta $
in the family $\left\{ \tilde{p}(x_{1}-\theta )\dots \tilde{p}(x_{n}-\theta
)\right\} ),$ see e.g. \cite{ru})$.$ By using the analog of (\ref{q*}), one
shows that for $\varphi \in {\cal H}_{\ast }^{(n)}$
\[
Q_{\ast }^{(n)}\varphi (\lambda _{1},\dots ,\lambda _{n})=i\frac{c^{\prime
}(\lambda )}{\sqrt{p^{(n)}(\lambda )}}\psi (\lambda _{1})\dots \psi (\lambda
_{n}),
\]
provided $c(\lambda )$ is absolutely continuous and $c^{\prime }(\lambda )$
is square integrable. We have
\[
c^{\prime }(\lambda )=\frac{1}{n}\sum_{j=1}^{n}\frac{\partial }{\partial
\lambda _{j}}c(\lambda _{1}+\dots +\lambda _{n}),
\]
hence
\begin{eqnarray*}
Q_{\ast }^{(n)}\varphi (\lambda _{1},\dots ,\lambda _{n}) &=&\frac{
i\sum_{j=1}^{n}\frac{\partial }{\partial \lambda _{j}}}{n}\left[ \frac{
c(\lambda _{1}+\dots +\lambda _{n})}{\sqrt{p^{(n)}(\lambda _{1}+\dots
+\lambda _{n})}}\psi (\lambda _{1})\dots \psi (\lambda _{n})\right] \\
&&-c(\lambda _{1}+\dots +\lambda _{n})\frac{i\sum_{j=1}^{n}\frac{\partial }{
\partial \lambda _{j}}}{n}\left[ \frac{\psi (\lambda _{1})\dots \psi
(\lambda _{n})}{\sqrt{p^{(n)}(\lambda _{1}+\dots +\lambda _{n})}}\right] .
\end{eqnarray*}
Taking into account that $\psi (\lambda _{j})=\sqrt{p(\lambda _{j})}{\rm e}
^{i\alpha (\lambda _{j})}$ and performing differentiation, we obtain after
some transformations
\begin{eqnarray*}
Q_{\ast }^{(n)}\varphi (\lambda _{1},\dots ,\lambda _{n}) &=&\frac{
\sum_{j=1}^{n}\left( i\frac{\partial }{\partial \lambda _{j}}+\alpha
^{\prime }(\lambda _{j})\right) }{n}\varphi (\lambda _{1},\dots ,\lambda
_{n}) \\
&&+\frac{i}{2}F(\lambda _{1},\dots ,\lambda _{n})\varphi (\lambda _{1},\dots
,\lambda _{n}),
\end{eqnarray*}
where
\[
F(\lambda _{1},\dots ,\lambda _{n})=\frac{[p^{(n)}(\lambda )]^{\prime }}{
p^{(n)}(\lambda )}-\frac{1}{n}\sum_{j=1}^{n}\frac{p^{\prime }(\lambda _{j})}{
p(\lambda _{j})}
\]
is a real function, or, briefly,
\[
Q_{\ast }^{(n)}\varphi =\frac{1}{n}\left( Q_{\ast }\otimes \dots \otimes
I+\dots +I\otimes \dots \otimes Q_{\ast }\right) \varphi +\frac{i}{2}
F\varphi ,\quad \varphi \in {\cal H}_{\ast }^{(n)}.
\]
Taking inner product with $\varphi ,$ and noticing that both $Q_{\ast
}^{(n)} $ and
\[
\frac{1}{n}\left( Q_{\ast }\otimes \dots \otimes I+\dots +I\otimes \dots
\otimes Q_{\ast }\right)
\]
are selfadjoint, we have $\langle \varphi |F\varphi \rangle =0,$ $\varphi
\in {\cal H}_{\ast }^{(n)},$ whence (\ref{sum}) follows.

Let us also show directly that $\langle \varphi |F\varphi \rangle =0,$ if $
\varphi $ is given by (\ref{phi}). We have
\begin{eqnarray}
\langle \varphi |F\varphi \rangle &=&\int \left\{ \frac{[p^{(n)}(\lambda
)]^{\prime }}{p^{(n)}(\lambda )}-\frac{1}{n}\sum_{j=1}^{n}\frac{p^{\prime
}(\lambda _{j})}{p(\lambda _{j})}\right\} \frac{|c(\lambda )|^{2}}{
p^{(n)}(\lambda )}p(\lambda _{1})\dots p(\lambda _{n})\,d\lambda _{1}\dots
d\lambda _{n}  \label{F} \\
&=&\int d\lambda \frac{|c(\lambda )|^{2}}{p^{(n)}(\lambda )}\int_{\Gamma
(\lambda )}\left\{ \frac{[p^{(n)}(\lambda )]^{\prime }p(\lambda _{1})\dots
p(\lambda _{n})}{p^{(n)}(\lambda )}-\frac{1}{n}\sum_{j=1}^{n}\frac{\partial
}{\partial \lambda _{j}}p(\lambda _{1})\dots p(\lambda _{n})\right\}
d^{n-1}\sigma .  \nonumber
\end{eqnarray}
But
\begin{eqnarray}
p^{(n)}(\lambda ) &=&\int_{\Gamma (\lambda )}p(\lambda _{1})\dots p(\lambda
_{n})d^{n-1}\sigma  \label{p(n)} \\
&=&p(\lambda )^{\ast n}=\int p(\lambda _{1})\dots p(\lambda _{n-1})p(\lambda
-\lambda _{1}-\dots -\lambda _{n-1})d\lambda _{1}\dots d\lambda _{n-1}.
\nonumber
\end{eqnarray}
Differentiating the last equality with respect to $\lambda ,$ we obtain
\begin{eqnarray*}
\lbrack p^{(n)}(\lambda )]^{\prime } &=&\int p(\lambda _{1})\dots p(\lambda
_{n-1})p^{\prime }(\lambda -\lambda _{1}-\dots -\lambda _{n-1})d\lambda
_{1}\dots d\lambda _{n-1} \\
&=&\int_{\Gamma (\lambda )}\frac{\partial }{\partial \lambda _{n}}p(\lambda
_{1})\dots p(\lambda _{n})d^{n-1}\sigma ,
\end{eqnarray*}
and similarly for all $\lambda _{j}.$ Hence
\begin{equation}
\lbrack p^{(n)}(\lambda )]^{\prime }=\frac{1}{n}\int_{\Gamma (\lambda
)}\sum_{j=1}^{n}\frac{\partial }{\partial \lambda _{j}}p(\lambda _{1})\dots
p(\lambda _{n})d^{n-1}\sigma .  \label{p'n}
\end{equation}
Taking into account (\ref{p(n)}), (\ref{p'n}) shows that the inner integral
in (\ref{F}) is equal to zero. $\Box $

\section{The Limit Theorem}

In this Section we impose the regularity assumption:

(A) $\psi \in {\cal D}(A)${\it , or, equivalently, }$\left\Vert A\psi
\right\Vert ^{2}=\int \lambda ^{2}p(\lambda )d\lambda <\infty ,${\it \ where
}$p(\lambda )=\left\Vert \psi (\lambda )\right\Vert ^{2}${\it \ is the
probability density of observable }$A${\it \ in the state }$S.${\it \
Without loss of generality we assume }${\sf E}_{\psi }(A)=\langle \psi
|A\psi \rangle =\int \lambda p(\lambda )d\lambda =0,${\it \ then }${\sf D}
_{S}(A)=\int \lambda ^{2}p(\lambda )d\lambda $ {\it \ is the variance of }$
A. $

The uncertainty relation (\ref{unc}) together with (\ref{H2.3.13}) imply the
lower bound for the variance of the optimal covariant estimate
\[
{\sf D}_{S^{(n)}}^{\ast }=\int_{{\bf R}}\left( \frac{d}{d\lambda }\sqrt{
p(\lambda )^{\ast n}}\right) ^{2}d\lambda \geq \frac{1}{4{\sf D}
_{S^{(n)}}(A^{(n)})}=\frac{1}{4n{\sf D}_{S}(A)},
\]
with equality attained if and only if $p(\lambda )$ is the Gaussian density.
The quantity ${\sf D}_{S^{(n)}}^{\ast }$ represents the accessible minimum
of variances of arbitrary (in general, entangled) estimates of the shift
parameter, based on $n$ independent observations.

By a variant of local central limit theorem \cite{pro},
\[
p_{n}(\lambda ):=\sqrt{n}\sigma p(\sqrt{n}\sigma \lambda )^{\ast
n}\rightarrow p_{0}(\lambda ):=\frac{1}{\sqrt{2\pi }}{\rm e}^{-\lambda
^{2}/2}
\]
in the sense of $L^{1}.$ Hence
\[
\int \left| \sqrt{p_{0}(\lambda )}-\sqrt{p_{n}(\lambda )}\right|
^{2}d\lambda \leq \int \left| p_{0}(\lambda )-p_{n}(\lambda )\right|
d\lambda \rightarrow 0.
\]
Therefore the Fourier transform of $\sqrt{p_{n}(\lambda )}$ converges to
that of $\sqrt{p_{0}(\lambda )}$ in $L^{2},$ and the probability density of
the optimal covariant observable
\[
p_{S^{(n)}}^{\ast }(x)=\frac{1}{2\pi }\left| \int {\rm e}^{-ix\lambda }\sqrt{
p(\lambda )^{\ast n}}d\lambda \right| ^{2}
\]
satisfies the local limit theorem
\[
\int \left| \sqrt{\frac{2}{\pi }}{\rm e}^{-2x^{2}}-\frac{1}{\sqrt{n}\sigma }
p_{S^{(n)}}^{\ast }\left( \frac{x}{\sqrt{n}\sigma }\right) \right|
dx\rightarrow 0.
\]
Thus, under the assumption (A) the distribution of the optimal covariant
observable $M_{\ast }^{(n)}$ in the state $S_{\theta }$ is asymptotically
normal with parameters ($\theta ,\frac{1}{4n\sigma ^{2}}$). In this sense
the bound of the uncertainty relation is asymptotically attainable.
Moreover, by using the main result of \cite{ba} one has the asymptotic
efficiency:

{\bf Theorem 3}. {\it If  }$p(\lambda )${\it \ is weakly differentiable and }
${\sf D}_{S^{(n)}}^{\ast }$ is {\it finite for some }$n${\it , then}
\[
\lim_{n\rightarrow \infty }n{\sf D}_{S^{(n)}}^{\ast }=\frac{1}{4{\sf D}
_{S}(A)}.
\]

\section{Semiclassical Estimation}

Consider now the estimation strategy when the optimal covariant quantum
estimate $M_{\ast }^{(1)}(dx)$ is found for every of the $n$ components in
the tensor product ${\cal H}^{\otimes n}$, and then the classical estimation
based on the obtained $n$ outcomes is made. The probability density of
observable $M_{\ast }^{(1)}(dx)$ in the state $S_{\theta }$ is
\[
\tilde{p}_{\theta }(x)=\frac{1}{2\pi }\left| \psi (x-\theta )\right| ^{2},
\]
where $\psi (x)=\int {\rm e}^{i\lambda x}\left\| \psi (\lambda )\right\|
d\lambda .$ Under the assumption (A), $\psi (x)$ and hence $\tilde{p}
_{\theta }(x)$ is differentiable, and for every unbiased estimate over $n$
observations the classical Cram\'{e}r-Rao inequality holds:
\[
{\sf D}_{n}\geq \left[ n\int \frac{|\tilde{p}_{\theta }^{^{\prime }}(x)|^{2}
}{\tilde{p}_{\theta }(x)}dx\right] ^{-1}.
\]
Notice that
\[
\int \frac{|\tilde{p}_{\theta }^{^{\prime }}(x)|^{2}}{\tilde{p}_{\theta }(x)}
dx=4\int \frac{\left| {\rm Re}\,\bar{\psi}^{^{\prime }}(x)\psi (x)\right|
^{2}}{\left| \psi (x)\right| ^{2}}\frac{dx}{2\pi }=4\int \left[ \left| \psi
(x)\right| ^{\prime }\right] ^{2}\frac{dx}{2\pi }.
\]
One the other hand,
\[
\int \lambda ^{2}\left\| \psi (\lambda )\right\| ^{2}d\lambda =\int \left|
\psi ^{^{\prime }}(x)\right| ^{2}\frac{dx}{2\pi }.
\]
Comparing this with the asymptotically attainable quantum bound of the
uncertainty relation, we have
\begin{eqnarray*}
\int \left[ \left| \psi (x)\right| ^{\prime }\right] ^{2}\frac{dx}{2\pi }
&<&\int \left| \psi ^{^{\prime }}(x)\right| ^{2}\frac{dx}{2\pi } \\
&=&\int \left[ \left| \psi (x)\right| ^{\prime }\right] ^{2}\frac{dx}{2\pi }
+\int \left| \beta ^{^{\prime }}(x)\psi (x)\right| ^{2}\frac{dx}{2\pi },
\end{eqnarray*}
if only $\beta (x):=\arg \psi (x)\neq {\rm const}$. Thus, under this
condition, for $n$ large enough
\[
n\min {\sf D}_{n}>n{\sf D}_{S^{(n)}}^{\ast },
\]
where the minimum is over all unbiased classical estimates using unentangled
quantum observables, which demonstrates superiority of the entangled quantum
estimation.

\section{Irregular Case}

\bigskip We now consider an instance of the Example where the regularity
assumption (A) does not hold. Let $S=|\psi \rangle \langle \psi |$ with
\[
\psi (\lambda )=\frac{1}{\sqrt{\pi a}}\frac{\sin a\lambda }{\lambda },
\]
so that ${\sf D}_{S}(A)=\infty .$ In the coordinate representation this
corresponds to the rectangular function
\[
\tilde{\psi}(x)=\int {\rm e}^{-i\lambda x}\psi (\lambda )d\lambda =\left\{
\begin{array}{c}
\sqrt{\frac{\pi }{a}},\quad {\rm if}\,x\in \lbrack -a,a]; \\
0,\quad {\rm if}\,x\notin \lbrack -a,a],
\end{array}
\right.
\]
that is, to the particle position uniformly distributed in $[-a,a],$ with
the probability density $\tilde{p}(x)=\frac{1}{2\pi }|\tilde{\psi}(x)|^{2}$
in ${\Bbb R}$. Unfortunately the probability density of the optimal
covariant observable $Q_{\ast }$ which is
\[
p^{\ast }(x)=\frac{1}{2\pi }\left\vert \int {\rm e}^{-i\lambda x}\left\vert
\psi (\lambda )\right\vert d\lambda \right\vert ^{2},
\]
cannot be found in explicit form, therefore we shall consider semiclassical
estimation based on unmodified position observable $Q$ having the
probability density $\tilde{p}(x-\theta )$.

Turning to the case of $n$ observations, we denote
\[
Q_{1}=Q\otimes \dots \otimes I,\,\dots ,\,Q_{n}=I\otimes \dots \otimes Q.
\]
As is well known (see e. g. \cite{kra}, n. 28.6), the Pitman estimate (\ref
{pit}) in the case of the rectangular probability density $\tilde{p}(x)$ has
the form
\[
\theta _{\ast }=\frac{1}{2}\left[ \min \left( Q_{1},\dots ,Q_{n}\right)
+\max \left( Q_{1},\dots ,Q_{n}\right) \right] .
\]
Its variance is equal to
\[
{\sf D}_{S}\left( \theta _{\ast }\right) =\frac{2a^{2}}{(n+1)(n+2)}\sim
\frac{2a^{2}}{n^{2}},
\]
which shows faster decay than $1/n$ characteristic to the regular case. The
rectangular density is nondifferentiable violating the regularity assumption
and making possible more efficient estimation than one which would follow
from the Cram\'{e}r-Rao bound (inapplicable in this case). Moreover,
denoting by $p_{n}(x)$ the probability density of $\theta _{\ast },$ one has
the limit law \cite{kra}
\begin{equation}
\lim_{n\rightarrow \infty }n^{-1}p_{n}(x/n)=\frac{1}{2a}{\rm e}^{-\frac{|x|}{
a}}.  \label{sub}
\end{equation}

Now we shall find the asymptotics of the optimal quantum covariant estimate
( \ref{sum}). Denoting by
\[
p(\lambda )=\left| \psi (\lambda )\right| ^{2}=\frac{1}{\pi a}\left( \frac{
\sin a\lambda }{\lambda }\right) ^{2}
\]
the probability density of $A,$ and by $f(x)=\int {\rm e}^{-i\lambda
x}p(\lambda )d\lambda ,$ we have
\[
f(x)=\left\{
\begin{array}{c}
1-\frac{|x|}{2a},\quad {\rm if}\,x\in \lbrack -2a,2a]; \\
0,\quad {\rm if}\,x\notin \lbrack -a,a].
\end{array}
\right.
\]
Thus
\[
p(\lambda )^{\ast n}=\frac{1}{2\pi }\int {\rm e}^{i\lambda x}f(x)^{n}dx,
\]
whence
\begin{eqnarray*}
\lim_{n\rightarrow \infty }np_{n}(n\lambda )^{\ast n} &=&\frac{1}{2\pi }
\lim_{n\rightarrow \infty }\int {\rm e}^{i\lambda x}f(x/n)^{n}dx \\
&=&\frac{1}{2\pi }\int \exp \left( i\lambda x-\frac{|x|}{2a}\right) dx=\frac{
1}{2\pi a}\frac{1}{\lambda ^{2}+(2a)^{-2}}.
\end{eqnarray*}
Taking into account that
\[
\int {\rm e}^{-i\lambda x}\frac{1}{\sqrt{\lambda ^{2}+(2a)^{-2}}}d\lambda
=2K_{0}\left( \frac{\left| x\right| }{2a}\right) ,
\]
where $K_{0}$ is the Macdonald (modified Bessel) function, and by using the
limit law, we can show (see Appendix) that the renormalized probability
density of the optimal observable
\[
p_{S^{(n)}}^{\ast }\left( x\right) =\frac{1}{2\pi }\left| \int {\rm e}
^{-i\lambda x}\sqrt{p(\lambda )^{\ast n}}d\lambda \right| ^{2}
\]
obeys the limit law
\begin{equation}
\lim_{n\to\infty}\int \left| \frac{1}{n}p_{S^{(n)}}^{\ast }\left( \frac{x}{n}
\right) -\frac{2 }{\pi a}\left| K_{0}\left( \frac{\left| x\right| }{2a}
\right) \right| ^{2}\right| dx=0.  \label{l1}
\end{equation}
With some more effort we can show (see Appendix) that its variance satisfies
\begin{eqnarray}
\lim_{n\rightarrow \infty }n^{2}{\sf D}_{S^{(n)}}^{\ast }
&=&\lim_{n\rightarrow \infty }\int_{{\bf R}}\left( \frac{d}{d\lambda }\sqrt{
np(n\lambda )^{\ast n}}\right) ^{2}d\lambda  \label{var} \\
&=&\frac{1}{2\pi a}\int_{{\bf R}}\left( \frac{d}{d\lambda }\sqrt{\frac{1}{
\lambda ^{2}+(2a)^{-2}}}\right) ^{2}d\lambda =\frac{a^{2}}{2},  \nonumber
\end{eqnarray}
which is four times less than for the semiclassical estimate $\theta _{\ast
}.$ Moreover, $K_{0}\left( x\right) \sim \sqrt{\frac{\pi }{2x}}{\rm e}^{-x}$
for large positive $x,$ whence
\[
\frac{2}{\pi a}\left| K_{0}\left( \frac{\left| x\right| }{2a}\right) \right|
^{2}\sim \frac{2}{\left| x\right| }{\rm e}^{-\frac{|x|}{a}},
\]
which shows that the tails of the asymptotic distribution of the optimal
estimate has somewhat faster decay than (\ref{sub}).

\section{Appendix}

\bigskip Proof of (\ref{var}). Denoting $f_{0}(x)={\rm e}^{-\left| x\right|
/2a},$ $p_{0}(\lambda )=\frac{1}{2\pi a}\frac{1}{\lambda ^{2}+(2a)^{-2}},$
\[
\Delta f_{n}(x)=f_{0}(x)-f(x/n)^{n};\quad \Delta p_{n}(\lambda
)=p_{0}(\lambda )-np(n\lambda )^{\ast n},
\]
we have

\begin{equation}
\Delta p_{n}(\lambda )=\frac{1}{2\pi }\int {\rm e}^{i\lambda x}\Delta
f_{n}(x)dx.  \label{fur}
\end{equation}
Now we observe that
\[
\lim_{n\rightarrow \infty }\int_{x\neq 0}\left| \Delta f_{n}(x)^{(k)}\right|
\left| x\right| ^{l}dx=0;\quad k,l=0,1,\dots .
\]
From (\ref{fur}) it follows in particular that
\begin{equation}
\max_{\lambda }\left| \Delta p_{n}(\lambda )\right| \leq \frac{1}{2\pi }\int
\left| \Delta f_{n}(x)\right| dx\rightarrow 0,\quad {\rm as}\,n\rightarrow
\infty .  \label{max}
\end{equation}
Let us also estimate the tails of $\Delta p_{n}(\lambda ).$ Taking into
account that $\Delta f_{n}(0)=\left( \Delta f_{n}(0)\right) ^{\prime }=0,$
and making twice integration by parts in (\ref{fur}), we obtain
\begin{equation}
\left| \Delta p_{n}(\lambda )\right| \leq \frac{1}{\lambda ^{2}}\int_{x\neq
0}\left| \Delta f_{n}(x)^{^{\prime \prime }}\right| dx=\frac{\varepsilon _{n}
}{\lambda ^{2}},  \label{dep}
\end{equation}
where $\lim_{n\rightarrow \infty }\varepsilon _{n}=0.$ In the same way we
obtain
\begin{equation}
\max_{\lambda }\left| \frac{d}{d\lambda }\Delta p_{n}(\lambda )\right| \leq
\frac{1}{2\pi }\int \left| x\Delta f_{n}(x)\right| dx\rightarrow 0,\quad
{\rm as}\,n\rightarrow \infty ,  \label{max1}
\end{equation}
and
\begin{equation}
\left| \frac{d}{d\lambda }\Delta p_{n}(\lambda )\right| \leq \frac{1}{\left|
\lambda \right| ^{3}}\int_{x\neq 0}\left| \left( x\Delta f_{n}(x)\right)
^{\prime \prime \prime }\right| dx=\frac{\varepsilon _{n}^{\prime }}{\left|
\lambda \right| ^{3}}.  \label{dep1}
\end{equation}
From (\ref{max}), (\ref{dep}) it follows that $\int \left| \Delta
p_{n}(\lambda )\right| d\lambda \rightarrow 0${\rm \ }as $n\rightarrow
\infty ,$ hence, arguing as before theorem 3, we have $\sqrt{np(n\lambda
)^{\ast n}}\rightarrow \sqrt{p_{0}(\lambda )}$ in $L^{2},$ implying (\ref{l1}
).

The estimates (\ref{max}), (\ref{dep}), (\ref{max1}), (\ref{dep1}) together
with
\[
p_{0}(\lambda )\asymp \frac{c}{1+\lambda ^{2}};\quad \left| \frac{d}{
d\lambda }p_{0}(\lambda )\right| \asymp \frac{c}{1+\left| \lambda \right|
^{3}}
\]
imply
\[
\left| \left( \frac{d}{d\lambda }\sqrt{p_{0}(\lambda )}\right) ^{2}-\left(
\frac{d}{d\lambda }\sqrt{p_{n}(\lambda )}\right) ^{2}\right| =\frac{1}{4}
\left| \frac{p_{0}^{\prime }(\lambda )^{2}}{p_{0}(\lambda )}-\frac{
p_{n}^{\prime }(\lambda )^{2}}{p_{n}(\lambda )}\right| \leq \frac{
\varepsilon _{n}^{\prime \prime }}{1+\lambda ^{2}},
\]
whence (\ref{var}) follows.

{\bf Acknowledgments.} The author acknowledges illuminating discussions with
participants of the Workshop ``Quantum Measurement and Quantum Stochastics''
held at Maphysto, University of Aarhus, Denmark, August 7-12, 2003, where
the initial version of this paper was presented. P. Harremoes pointed out a
remarkable connection between the Theorem 3 and the Fisher information
approach to the central limit theorem in classical probability. The recent
development of this approach (going back to Yu. V. Linnik's 1959 paper) in
the reference \cite{ba} provided to us by P. Harremoes enables to give a
proof for Theorem 3 under the minimal regularity assumptions. Other comment
concerns our restriction to covariant estimates which appears to be
essential. As noticed by M. Hayashi and K. Matsumoto, in the broader context
of locally unbiased estimates entangled observables do not give advantage in
one-parameter families of quantum states, which can be shown by using
adaptive estimation trick due to R. Gill \cite{ma}. The author is grateful
to these participants as well as to the organizers of the Workshop.

This work was partially supported by INTAS grant 00-738.


\begin{thebibliography}{99}
\bibitem{4*}  B a r n d o r f f - N i e l s e n \ O. E., G i l l \ R. D., J
u p p \ P. E. On quantum statistical inference. J. Royal Statist. Soc. B,
{\bf 65} , 1-31, 2003.

\bibitem{kra}  C r a m \'{e} r \ H. Mathematical Methods of Statistics.
Stockholm, 1946.

\bibitem{7*}  D'A r i a n o G. M. Homodyning as universal detection. In:
Quantum Communication, Computing and Measurement. Eds. Hirota O., Holevo A.
S., Caves C. M. New York: Plenum Press, pp. 253-264, 1997; LANL e-print
quant-ph/9701011.

\bibitem{der} D e r k a \ A., B u z e k \ V., E k e r t \ A. Universal
algorithm for optimal estimation of quantum states from finite ensembles; LANL e-print
quant-ph/9707028.


\bibitem{ma}  G i l l \ R. D., M a s s a r \ S. State estimation for large
ensembles. Phys. Rev. A {\bf 61}, 042312/1-16, 2000; LANL e-print
quant-ph/9902063.

\bibitem{dar}  G i l l \ R., G u t a \ M. I. An invitation to quantum
tomography. LANL e-print quant-ph/0303020.

\bibitem{hay}  H a y a s h i \ M. Asymptotic estimation theory for a finite
dimensional pure state model. J. Phys. A {\bf 31}, 4633-4655, 1998.


\bibitem{hel76}  H e l s t r o m \ C.W. Quantum Detection and Estimation
Theory, Acad. Press, New York, 1976. Russian translation: Moscow Mir, 344 pp
(1979)

\bibitem{hol80}  H o l e v o \ A.S. Probabilistic and Statistical Aspects of
Quantum Theory, Moscow, Nauka, 1980. English translation: North Holland,
Amsterdam, (1982)

\bibitem{hol83}  H o l e v o \ A. S. Bounds for generalized uncertainty of
shift parameter. Lect. Notes Math., {\bf 1021}, 243-251, 1983.

\bibitem{hol2}  H o l e v o \ A. S. Generalized imprimitivity systems for
Abelian groups, Izv VUZ. Matematika, N2, 49-71, 1983. English translation:
J. Soviet Math., 53-80.

\bibitem{2*}  H o l e v o \ A. S. Statistical Structure of Quantum Theory.
Lect. Notes Phys. {\bf m67}, Springer-Verlag: New York-Heidelberg-Berlin,
2001.

\bibitem{ba}  J o h n s o n \ O., B a r r o n \ A. Fisher information
inequalities and the central limit theorem. Preprint, 2003.

\bibitem{key}  K e y l \ M., W e r n e r \ R. F. Estimating the spectrum of
a density operator. Phys. Rev. A, {\bf 64}, no.5, 052311, 2001.

\bibitem{ma1}  M a s s a r \ S., P o p e s c u \ S. Optimal extraction of
information from finite quantum ensembles. Phys. Rev. Lett. {\bf 74},
1259-1263, 1995.

\bibitem{pro}  P r o k h o r o v \ Yu. V. Local theorem for densities.
Doklady AN SSSR. {\bf 83}, 797-800, 1952. (rus)

\bibitem{ru}  R u k h i n \ A. L. Some statistical and probabilistic
problems on groups. Proc. Steklov Mathematical Institute. {\bf 111}, 52-109,
1970. (rus)
\end{thebibliography}
\end{document}